\documentclass[aps,prl,reprint,showpacs]{revtex4-1}
\usepackage{graphicx}
\usepackage{amsmath}

\begin{document}

\title{Detection of early cluster formation in soft matter:
 an entropic imprint}
\date{\today}
\author{Jean-Marc Bomont}
\email[Corresponding author, e-mail: ]{bomont@univ-metz.fr\\}
\author{Jean-Louis Bretonnet}
 \affiliation{
Laboratoire de Physique des Milieux Denses, Universit\'{e} de Metz
1 Bd. Arago, 57\,078 Metz -- France}
\author{Dino Costa}
\affiliation{Dipartimento di Fisica and
CNISM (Consorzio Nazionale Interuniversitario di Struttura della Materia)\\
Universit\`a degli Studi di  Messina, 
Viale F. Stagno d'Alcontres 31, 98166 Messina -- Italy
}

\begin{abstract}
We study model protein solutions and
colloidal suspensions in the
temperature range whereupon the nature of the
system changes from a homogeneous fluid to a ``cluster fluid''.
It is commonly assumed~---~as deduced by the behavior of
the structural correlations~---~that this microphase separation
sets as a continuous process. We challenge such assumption and demonstrate
that the entropy shows a discontinuous behavior across a
well-defined temperature threshold, providing a neat
signature of the modifications occurring in the fluid.
All predictions, obtained in the framework of a 
refined theory of the liquid state,
are systematically assessed against extensive Monte 
Carlo simulations. We emphasize
the broad generality of our conclusions, and
the importance of accurate theoretical treatments
to characterize the thermodynamic properties of soft materials.
\end{abstract}

\pacs{05.20.Jj, 61.20.Gy, 02.30.Rz, 61.20.Ja}

\maketitle

Recent small-angle scattering and
confocal microscopy studies of globular protein solutions 
and colloid-polymer 
mixtures
identified, upon appropriate thermodynamic conditions, 
a secondary peak in
the structure factor $S(q)$ at a wave vector $q$ smaller than that of
the main diffraction peak~\cite{Stradner:04,Baglioni:04}. 
Such a peculiar
feature differs significantly from what is usually observed in
``simple liquids'', because the appearance of
the peak is attributed to structural modifications inherent to the
formation of 
stable clusters 
in the homogeneous fluid phase.
Early theoretical studies on aggregation 
in solution date back to the 40's work of Debye~\cite{Debye:49}
and focused on the thermodynamic  properties of phenomenological 
models.
Progress
elucidating the mechanisms
underlying the aggregation processes was made only recently;
in particular,
the small-$q$ peak is now interpreted
in terms of enhanced density fluctuations which do not
evolve into a fully developed macroscopic phase 
separation~\cite{Sear:99,Pini:00,Groenewold:04,Imperio:04}.
Investigations in this area
attracted increasing attention because they may shed light on
aggregation processes that can precede arrested states,
see e.g.~\cite{Baglioni:04,Wu:04,Sciortino:04,%
Campbell:05,Decandia:06,Lu:08,Toledano:09}; 
more generally, the control of
aggregation constitutes a crucial step in disparate
realms of science
and technology.

Theory and experiments showed that
the growth of the small-$q$ peak
sets as a continuous process; moreover,
the early formation of aggregates is expected to be well
underway before such a feature
clearly displays.
 Hence, a main issue concerns
how to unequivocally
identify the temperature
threshold whereupon
the nature of the system changes from a homogeneous
fluid to a ``cluster fluid'',
as well as to characterize
from a thermodynamic viewpoint
the ``quality'' of the changes occurring in the fluid.
This Letter provides an answer to such issues,
identifying
a discontinuity
in the entropy (and other thermodynamic potentials) 
across the temperature 
whereupon the aggregation process starts to take place.
Such discontinuous features were normally observed
in a variety of colloidal models, in the presence  of 
a phase transition
either between a homogeneous
fluid (vapor or liquid) and a ``modulated phase'' (e.g. clusters,
stripes or bubbles),
or between different 
modulated phases, see e.g~\cite{Sear:99,Imperio:04,%
Tarzia:06,Decandia:06,Archer:07,Archer:08}.
Here we document~---~by theory and 
simulations~---~that a discontinuous behavior emerges even
in the absence of a (macro)phase separation, 
as the system 
experiences only a transformation
between a normal homogeneous fluid and a fluid with
local inhomogeneities, characterized by the
formation of cluster aggregates. As we shall show,
the observed behavior
is consistent with early predictions~\cite{Debye:49}.

The minimal requirements a basic microscopic
interaction should possess in order to develop and stabilize
protein or colloid aggregates
seems to be 
an appropriate balance between short-range attraction (favoring the
cluster growth at low enough temperature) and long-range repulsion 
(preventing a complete phase separation, so to stabilize the
aggregate formation)~%
\cite{Stradner:04,Sear:99,Sciortino:04,brazovskii:75,%
Campbell:05,LIU2005,Broccio:06,Cardinaux:07,Chen:07}. 
The long-range repulsion is attributed to screened electrostatic
interactions, or to the presence of 
co-solutes in solution~\cite{Barrat}, whereas
short-range forces stem from several factors, including van der Waals
interactions, hydrophobic effects, depletion mechanisms~\cite%
{Lu:08,Tardieu:01,Louis:02,Bonn:09}.
Therefore, spherical 
particles interacting through 
a two-Yukawa (2Y) potential in the form
\begin{equation}
u(r)=\left\{
\begin{array}{cc}
+\infty & r<\sigma , \\[8pt]
 -\dfrac{K_{1}\, {\rm e}^{-Z_{1}(r -\sigma)}} {r/\sigma } 
 +\dfrac{K_{2}\, {\rm e}^{-Z_{2}(r-\sigma)}} {r/\sigma } & r\geq \sigma%
\end{array}%
\right.  \label{eq:pot}
\end{equation}%
(giving rise
to a short-range attraction followed by
a long-range repulsion through
the requirements $K_{1}>K_{2}$ and $Z_{1}>Z_{2}$)
should provide a reasonably flexible and generic 
model to investigate, from a microscopic
viewpoint,  the
cluster formation in protein and colloidal 
solutions.
More generally, 2Y systems~---~and similar models with
competing interactions~---~exhibit a rich variety of 
inhomogeneous fluid phases and intriguing phase
behaviors; as such they are lively studied both
theoretically and in terms of 
computer simulation (see~\cite{Sear:99,Archer:07,BomontCosta2010}
for a detailed bibliography).

In this Letter we provide a theoretical characterization, 
in terms of the 
entropy behavior, of the 
temperature threshold whereupon the homogeneous 2Y 
fluid changes its nature
 into a cluster fluid.
Entropy
is a well-recognized indicator of structural modifications occurring in
the fluid~\cite{Martynov,Giaquinta,JMB2003,Saitta:06,JMB2008,Lee:10};
moreover,
the interpretation of experimental results often resorted 
on the use of theoretical
approaches, represented for instance by 
integral equation theories (IET)~\cite{JMB2008,TOSL}
of the liquid state, see e.g.~\cite{LIU2005,Tardieu:01}.
Hereby, in order to determine the structural
properties of the fluid we have solved the Ornstein-Zernike equation coupled
with a refined closure relation enforcing the 
thermodynamic consistency at the level of structural 
properties~\cite{JMB2008}. 
On top of such first condition, 
a {\rm second thermodynamic consistency condition} 
is introduced,
involving the calculation of 
the chemical potential~\cite{JMB2003,JMB2007}.
We have
recently documented the accuracy of the whole scheme 
for the 2Y model~\cite{BomontCosta2010}.
All theoretical predictions have been gauged against
extensive Monte Carlo (MC) simulations~\cite{note-mc}.

The cluster formation
is analyzed 
as a function of the temperature
for three different ratios
$K_{2}/K_{1}$ of the repulsion over 
the attraction strength 
($K_{2}/K_{1}=0.01$, 0.05 and 0.1); 
the reduced temperature is defined
as $T^*=k_{\mathrm{B}}T/(K_{1}-K_{2})$, with $k_{\mathrm{B}}$ as the
Boltzmann constant.
We set $Z_1=10$, in order to have a $\approx 10$\%
ratio of the attraction range over the diameter $\sigma$
(to be used as length unit),
as in typical
globular protein solutions; $Z_2$ is fixed at 0.5, 
corresponding to a weak
ionic strength of order $I=2$\,mM, as obtained
with $\sigma\approx33/37$\AA, a typical diameter range
used in protein models~\cite{LIU2005}.
Our choice of numerical values
implies the absence of a
liquid-vapor phase separation for the 
model at issue~\cite{Pini:00}; moreover,
due to the relatively long-range character of
the repulsive contribution,
the system is expected  to
form~\cite{Sciortino:04}, for low enough temperatures,
a Wigner glass of clusters,
i.e. a disordered state of polydisperse clusters,
that do not percolate to give a gel.
As pointed out elsewhere~\cite{LIU2005}, 
when suitable conditions for the 
cluster formation are met~---~for 
a given choice of control parameters 
in Eq.~(\ref{eq:pot})~---~there is an
optimal packing fraction with a corresponding largest
intensity of the small-$q$ peak.
In our case we have estimated such
packing fraction  as $\phi \equiv \pi \rho \sigma ^{3}/6=20\%$ of the total volume,
where $\rho\sigma^3$ is the reduced number density.

\begin{figure}[!t]
\begin{center}
\includegraphics[angle=-90,width=8.0cm]{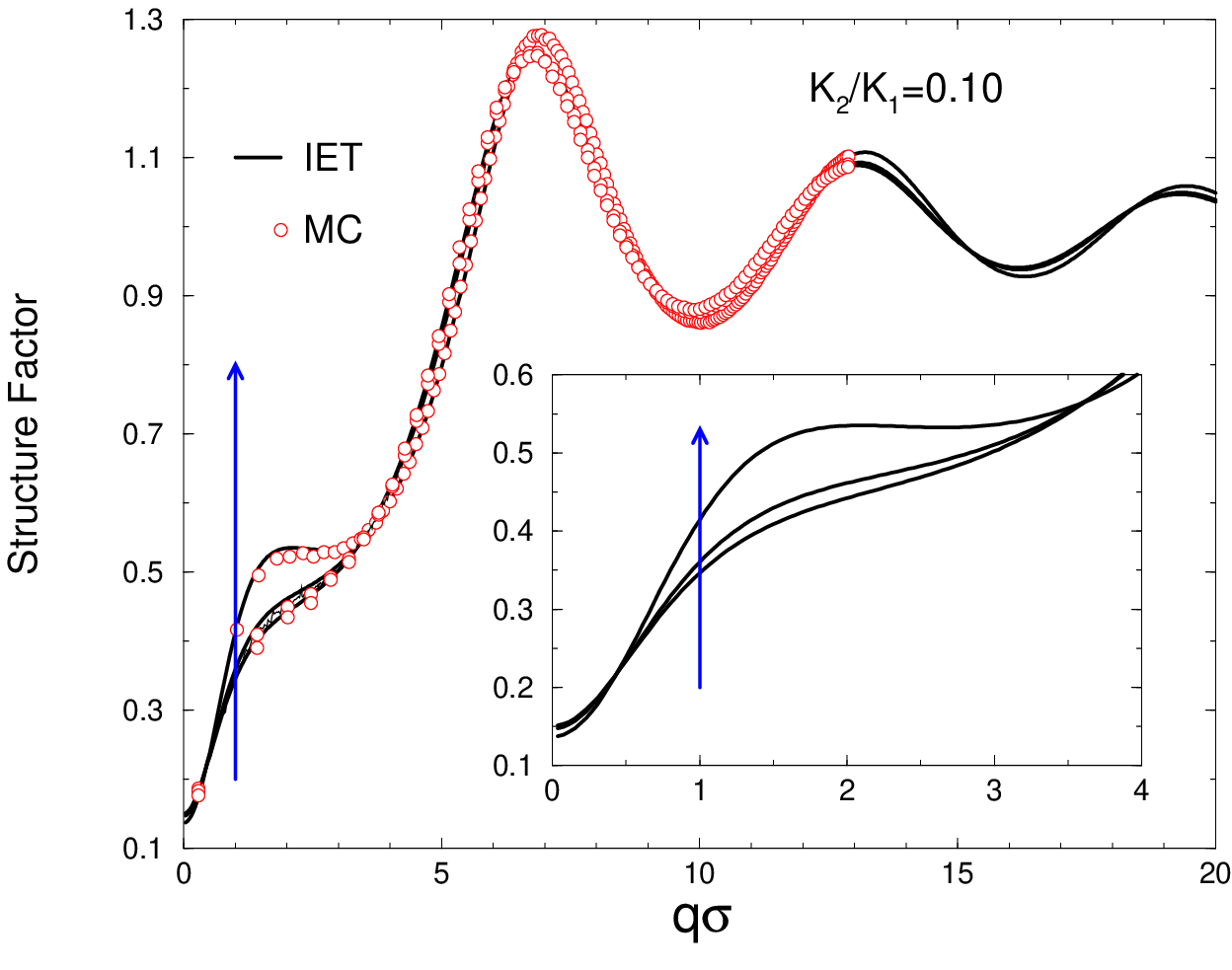}
\end{center}
\caption{
IET (lines) and MC (symbols)
structure factors of a 2Y fluid with $K_2/K_1=0.10$,
at a packing fraction $\phi=20$\% and
different temperatures ($T^*=0.74$, 0.687, and 0.55,
in the direction of the arrow).
Inset: magnification of the small-$q$ region
where the cluster peak develops.
}
\label{fig:sq}
\end{figure}

At the microscopic level, 
the aggregation process
may be regarded as the creation of bonds between close particles, with
two particles considered as bonded together as soon as their average
separation is less than a small fraction of 
the diameter $\sigma $ and stays almost
unchanged with decreasing temperature~\cite{Sciortino:04}.
At the structural level such process transpires 
from the behavior of the structure factor:
in Fig.~\ref{fig:sq} we document
the
progressive enhancement of the secondary peak at 
small-$q$ vectors 
(at $K_2/K_1=0.10$),
as the temperature decreases.
As alluded to, the almost continuous growth of such a
peak~---~initially manifesting 
as an inflection point without a local maximum, 
see also the inset of Fig.~\ref{fig:sq}~---~is considered
as a signal of well-formed clusters in the fluid.

At the macroscopic level, the Debye idealized ``two-state'' 
model~\cite{Debye:49}
assumes that,  at a given temperature, a substance
can exist in a solution
either as a collection of
monomers $A$, or as aggregates of $n$ monomers $A_n$; the 
equilibrium constant of the reaction $nA \to A_n$ is 
given by $\exp[-\beta\,\Delta\mu]$, 
where $\Delta\mu$ denotes the variation of the chemical 
potential when 
forming an aggregate out of $n$ monomers.
In other words, thermodynamics predicts that the appearance 
of clusters should 
be detected by a variation in $\mu$.
Intuitively, one may assume that,
in the homogeneous fluid,
the ``pre-existing form''
of a cluster fills 
a certain physical volume; then, at a given temperature,
such a volume suddenly shrinks, via the bonding processes illustrated above,
while the space available for 
the remaining particles, not involved in the
aggregation process, becomes larger. We may then expect
that at this peculiar stage the insertion
of any additional particle in the sample should become an easier task, in
comparison with previous conditions, and that the corresponding 
energy cost~---~defining exactly the excess chemical potential 
of the fluid~---~should therefore
decrease.

\begin{figure}[!t]
\begin{center}
\includegraphics[angle=-90,width=8.0cm]{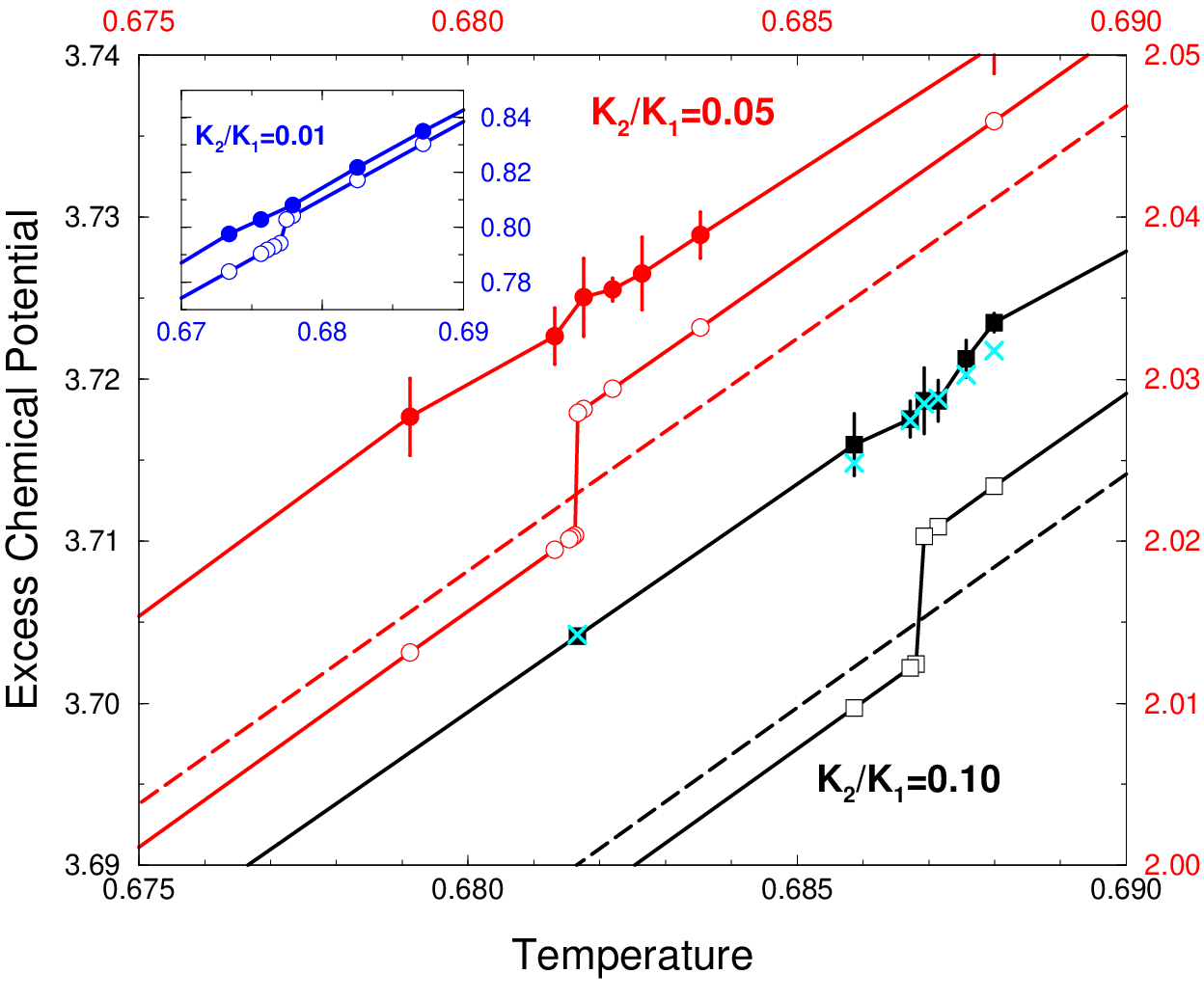}
\end{center}
\caption{IET (open symbols)
and MC (full symbols with error bars)
excess chemical potentials~\cite{note-mc}
for $K_2/K_1=0.10$ (squares), 0.05 (circles), 
and 0.01 (triangles, inset). MC data refer
to a sample of $N=4096$ particles; crosses are MC data for
$K_2/K_1=0.10$ and $N=13824$.
Predictions from the Kiselyov-Martynov scheme~\cite{Martynov}
at $K_2/K_1=0.05$ and 0.10
are also shown as dashed lines.
Follow the color code to identify the
horizontal and vertical scales.
Full lines are guides to the eye.
}
\label{fig:mu}
\end{figure}

Our theoretical predictions for the chemical potential, 
reported in Fig.~\ref{fig:mu}, present
the expected behavior:
for each $K_{2}/K_{1}$ ratio, the excess
chemical potential displays a
monotonic behavior as a function of the temperature until
a threshold temperature  $T^*_{\rm C}$ is reached, whereupon a 
discontinuous decrease
invariably occurs.
The overall behavior of the chemical potential
is confirmed by the corresponding Monte Carlo Widom test-particle 
calculations~\cite{note-mc} (to within $< 0.5\%$ discrepancy),
also reported in Fig.~\ref{fig:mu}.
As visible, the MC discontinuities  
practically occur at the same temperature predicted by IET,
even if they appear much 
weaker and broader 
than the theoretical ones 
(i.e. the differences in the MC chemical
potential across $T^*_{\rm C}$ are smaller and,
within the resolution and statistical uncertainties
of our simulations, we rather observe a rapid decrease
over a tiny temperature range).
MC data mostly refer to a sample
composed of $N=4096$ particles;
results with $N=13824$
at $K_2/K_1=0.10$ (also shown in Fig.~\ref{fig:mu}),
substantially confirm the trends observed for the 
smaller system size.
The discontinuities fall
in the same tiny temperature range, almost independently
on the ratio $K_2/K_1$; 
we may then expect that
$T^*_{\rm C}$ mostly depends on the density of the fluid.
Moreover, the strength of the discontinuity
decreases as the $K_2/K_1$ ratio is lowered,
consistently with the 
fact that
in these conditions the fluid behaves 
progressively (and strictly at $K_2=0$) 
as a ''simple'' one. 
As expected, $T^*_{\rm C}$
anticipates (i.e. is higher than) the temperature
whereupon the cluster  peak develops in $S(q)$: we obtain, in fact,
for the pairs $[K_2/K_1; T^*_{\rm C}]$ 
the IET values
$[0.01; 0.676]$, $[0.05; 0.682]$ and $[0.1; 0.687]$,
to be compared
 with our estimate for the early appearance of the cluster peak
$[0.01; \approx0.64]$, $[0.05; \approx0.58]$ and $[0.1; \approx0.55]$ 
(see also the inset in Fig.~\ref{fig:sq}). 

We have verified that the discontinuity  in the chemical potential
is equally well-predicted if the
second thermodynamic consistency condition is enforced
on top of different, equally  refined closure relations, 
giving accurate, self-consistent 
structural predictions.
On the other hand, in the absence
of the second thermodynamic consistency 
condition,
the whole
theoretical approach results
completely insensitive to the cluster formation; as an example
we report in Fig.~\ref{fig:mu} the flat response
of the excess chemical potential calculated
within the otherwise accurate Kiselyov and
Martynov scheme~\cite{Martynov}.
As discussed,
the discontinuity in the chemical
potential occurs in a narrow temperature range, and
manifests as a tiny jump of few percents of the total values:
previous simulations could have missed such feature
possibly because of a lack of initial knowledge, as
provided here by IET predictions.

\begin{figure}[!t]
\begin{center}
\includegraphics[angle=-90,width=7.9cm]{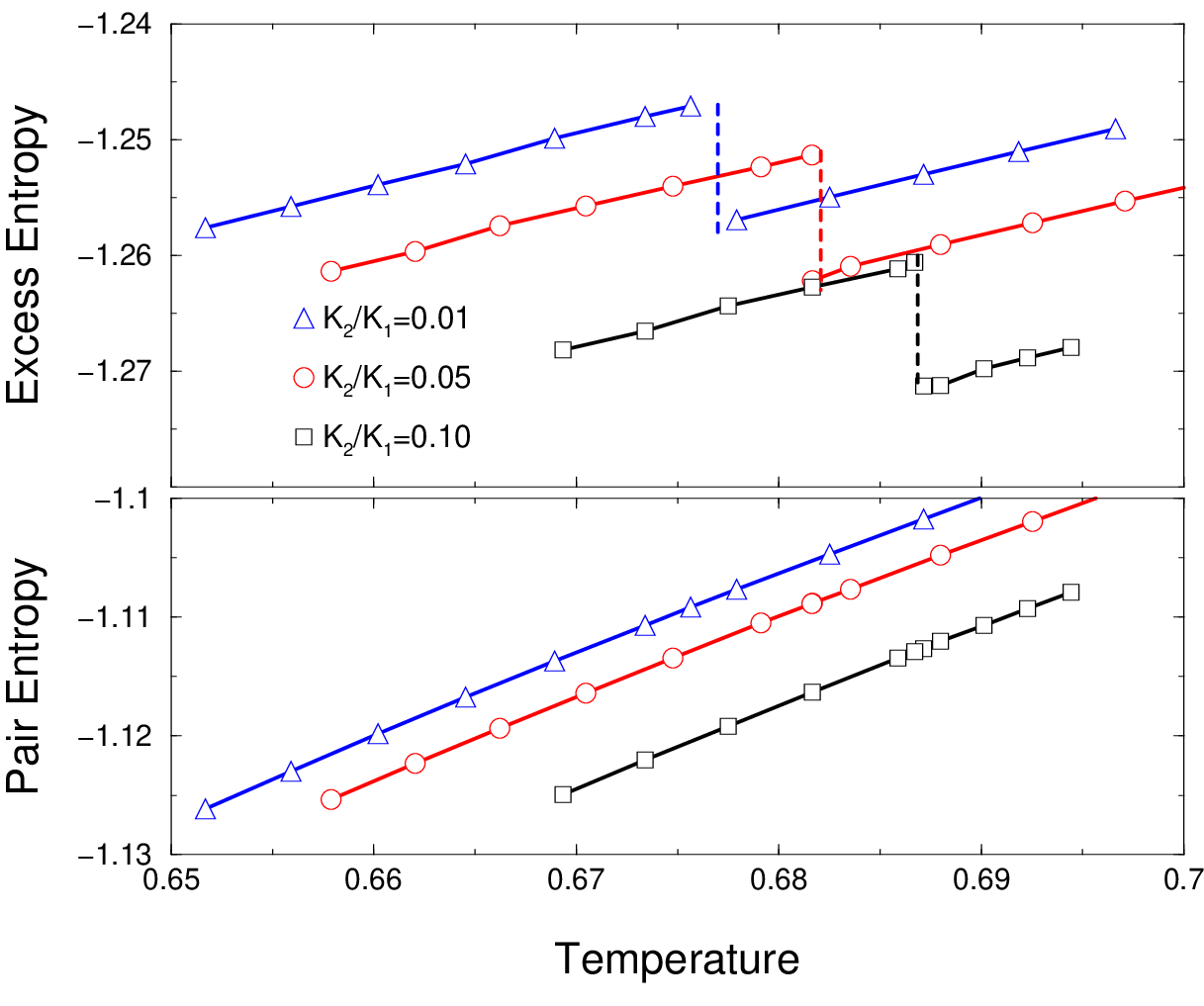}
\end{center}
\caption{IET excess (top) and pair (bottom) entropies.
The discontinuities
at $T^*_{\rm C}$ are indicated by the vertical dashed lines.
Full lines are guides to the eye.
}
\label{fig:entropy}
\end{figure}

We may now turn to the notion of order in the 2Y fluid, and examine
the consequences of the above 
calculated insertion energies on the total
entropy.
We know that the more the
temperature decreases, the more the system becomes ordered. Yet, may we
assume that at the same time
the entropy monotonically decreases?
As can be seen
from Fig.~\ref{fig:entropy},  the answer is negative
for the process at issue:
as far as $T^*$ is higher than $T^*_{\rm{C}}$,
bonds are not established, and the excess entropy $S^{\rm ex}$
normally decreases, indicating that the system gradually orders;
at the threshold $T^*_{\rm{C}}$
the nature of the fluid changes and the discontinuity in 
$\mu^{\mathrm{ex}}$ has a faithful counterpart in the behavior of 
$S^{\rm ex}$ (and hence in the 
behavior of the free energy, not reported here). 
The loss of available (configurational) space, due to
localization, 
is overcome by the overall gain of accessible 
(physical) space, 
and therefore the entropy discontinuously increases. 
Below $T^*_{\rm{C}}$, the entropy regains the normal behavior
and monotonically restarts to decrease.
As for MC results, a clear picture reveals in this case 
more problematic to emerge, 
essentially  because 
the statistical 
uncertainties in the calculation of the chemical potential, 
the pressure, and the internal energy cumulatively prevent 
a finely resolved determination of the entropy.
We may also analyze the discontinuous behavior of $S^{\rm ex}$
in the framework of the entropy multiparticle-correlation expansion
derived by Nettleton and Green in the late 50's~\cite{Nettleton}. 
In this framework,
the pair entropy $S_2$, given by a spatial integral of
(a functions of) the pair correlation function, has been often used as a
straightforward approximation for the whole entropy,
see e.g.~\cite{Sharma} and references therein. 
Such approximation
largely fails for the aggregation process at issue, 
as $S_2$ decreases
monotonically with the temperature (see Fig.~\ref{fig:entropy}): 
in fact the pair entropy,
``trivially'' reading the progressive local
structuring of the fluid~\cite{Giaquinta}, 
bears no trace of
the observed jump in 
$S^{\rm ex}$, that must be reflected in the higher-order terms of
the Nettleton and Green expansion.

To summarize, 
we have shown that a neat thermodynamic imprint~---~given by a discontinuity
in the entropy and other thermodynamic potentials as functions
of the temperature~---~characterizes the ``homogeneous to
cluster fluid'' microphase transformation in
two-Yukawa models.
The observed behavior 
is consistent with the predictions
of early phenomenological 
approaches~\cite{Debye:49}.
We have emphasized the key role
played in this study 
by an accurate statistical mechanics approach.

\begin{acknowledgments}
The authors warmfully thank C.~Caccamo, P.~V.~Giaquinta,
J.-P.~Hansen, A. Parola, S.~Prestipino and F. Sciortino for their
advice and helpful discussions.
The computer time provided by 
the P\^{o}le Messin de Mod\'elisation et Simulation 
and the Centro di Calcolo Elettronico 
dell'Universit\`a degli Studi di Messina 
``A.~Villari'' is 
gratefully acknowledged.
\end{acknowledgments}


\begin{thebibliography}{99}
\bibitem{note-mc}
Simulations have been carried out on a sample composed of $N=4096$
particles, enclosed in a cubic box with periodic
boundary conditions 
(corresponding to an interaction cutoff of $\approx 11\sigma$).
Simulations at $K_2/K_1=0.10$ and $N=13824$ have been 
done to detect any possible size dependence of our results.
We have calculated
the excess chemical potential through the Widom
test particle method~\cite{TOSL}.
Standard long-range corrections
have been added to all simulation data; the $O(1/N)$
corrections to the chemical potential~\cite{Smit:89}, as estimated 
from the accurate IET compressibilities, are
comparatively negligible.
Error bars are obtained by dividing the MC runs into 5/15 sub-blocks
of 200\,000 moves per particle each.
\end{thebibliography}
\end{document}